\documentclass[epj]{webofc}
\usepackage[utf8]{inputenc}
\usepackage[varg]{txfonts}   
\usepackage{booktabs}
\usepackage{xcolor}
\definecolor{darkred}{rgb}{0.4,0.0,0.0}
\definecolor{darkgreen}{rgb}{0.0,0.4,0.0}
\definecolor{darkblue}{rgb}{0.0,0.0,0.4}
\usepackage[bookmarks,linktocpage,colorlinks,
    linkcolor = darkred,
    urlcolor  = darkblue,
    citecolor = darkgreen]{hyperref}
\usepackage{amsmath}
\wocname{EPJ Web of Conferences}
\woctitle{Lattice2017}
\newcommand{\up}{\uparrow}
\newcommand{\down}{\downarrow}
\newcommand{\pf}{\mathrm{pf}}
\begin{document}
%
\selectlanguage{english}
\title{Lattice simulation with the Majorana positivity}
\author{%
\firstname{Arata} \lastname{Yamamoto}\inst{1}
\fnsep\thanks{Speaker, \email{arayamamoto@nt.phys.s.u-tokyo.ac.jp}}
\and
\firstname{Tomoya} \lastname{Hayata}\inst{2}
}
\institute{%
Department of Physics, The University of Tokyo, Tokyo 113-0033, Japan
\and
Department of Physics, Chuo University, Tokyo, 112-8551, Japan
}
\abstract{%
While the sign problem of the Dirac fermion is conditioned by the semi-positivity of a determinant, that of the Majorana fermion is conditioned by the semi-positivity of a Pfaffian.
We introduce one sufficient condition for the semi-positivity of a Pfaffian.
Based on the semi-positivity condition, we study an effective model of the Majorana fermion.
We also present the application to the Dirac fermion.
}
\maketitle

\section{Majorana positivity}

The sign problem is one of the most important subjects in computational physics.
There are many fermion systems with the sign problem, such as QCD with quark chemical potential and the repulsive Hubbard model.
There are also many fermion systems without the sign problem, such as QCD with isospin chemical potential and the attractive Hubbard model.
The study of such sign-problem-free systems is useful for us since it can be exactly performed by the Monte Carlo method.
How can we discover new sign-problem-free theory?

The conventional argument on the absence condition of the fermion sign problem is as follows.
When the action is written by the Dirac (or complex) fermion as
\begin{equation}
 S_\psi = \int d^dx \bar{\psi} D \psi
,
\end{equation}
the path integral is given by
\begin{equation}
\begin{split}
 Z 
&= \int D\bar{\psi} D\psi DA \ e^{ - S_\psi - S_A }
\\
&= \int DA \ \det D \ e^{-S_A}
.
\end{split}
\end{equation} 
If the determinant is semi-positive,
\begin{equation}
 \det D \ge 0
,
\end{equation}
the theory is sign-problem free.
Here we introduce unconventional argument.
When the action is given by the Majorana (or real) fermion as
\begin{equation}
 S_\Psi = \int d^dx \frac{1}{2} \Psi^\top P \Psi
,
\end{equation}
the path integral is given by
\begin{equation}
\begin{split}
 Z 
&= \int D\Psi DA \ e^{ - S_\Psi - S_A }
\\
&= \int DA \ \pf P \ e^{-S_A}
.
\end{split}
\end{equation}
The determinant $\det D$ is replaced by the Pfaffian $\pf P$.
The sign-problem-free condition becomes the semi-positivity of the Pfaffian
\begin{equation}
 \pf P \ge 0
.
\end{equation}
This argument based on the semi-positivity of a Pfaffian is called the Majorana positivity.
The Majorana positivity is applicable not only to Majorana fermions but also to Dirac fermions.
One Dirac fermion is decomposed into two Majorana fermions,
\begin{equation}
\psi = \frac{1}{\sqrt{2}}(\Psi_1 + i \Psi_2)
.
\end{equation}
The path integral is written by the two Majorana fermions and its sign-problem-free condition is given by the semi-positivity of their Pfaffian.
The Majorana positivity was originally proposed to find sign-problem-free fermion models in condensed matter physics \cite{Li:2014tla}.

The Pfaffian is defined for even-dimensional anti-symmetric matrices.
The general form is
\begin{equation}
P = 
\begin{pmatrix}
P_1 & iP_2\\
-iP_2^\top & P_3
\end{pmatrix} 
,
\end{equation}
where $P_1$ and $P_3$ are anti-symmetric block matrices.
We assume the block matrices $P_1$, $P_2$, and $P_3$ are even-dimensional although it is not essential \cite{Hayata:2017jdh}.
In general, the Majorana positivity condition is not unique.
We present one sufficient condition.
If both of
\begin{description}
\item[condition 1] $P_2$ is semi-positive
\item[condition 2] $P_3=-P_1^\dagger$ and $P_2 = P_2^\dagger$
\end{description}
are satisfied, the Pfaffian $\pf P$ is semi-positive definite \cite{Hayata:2017jdh}.
In this presentation, we will not show the mathematical proof but show the examples satisfying this Majorana positivity condition.
We use the lattice unit in all the following equations.

Before showing the examples, we briefly comment on simulation scheme.
When the Majorana positivity condition is satisfied, the simulation scheme is straightforward.
Because of $\pf P \ge 0$,
\begin{equation}
\begin{split}
 Z &= \int DA \  \pf P \ e^{-S_A}
\\
&= \int DA \ (\det P)^{\frac{1}{2}} \ e^{-S_A}
\\
&= \int DA \ (\det PP^\dagger)^{\frac{1}{4}} \ e^{-S_A}
.
\end{split}
\label{eqHMC}
\end{equation}
This is the standard form simulated by the Hybrid Monte Carlo method with pseudo-fermion fields.
Note that, if $\pf P$ is indefinite, Eq.~\eqref{eqHMC} does not hold true because of nontrivial phase factors.

\section{Majorana fermions in 1+1 dimensions}

The first example is the Majorana fermion.
One motivation to study the Majorana fermion is supersymmetry in particle physics \cite{Catterall:2009it}.
It is also studied in condensed matter physics, for example, in the context of the surface states of topological superconductors \cite{Hasan:2010xy}.

Let us consider the two-component Majorana fermion
\begin{equation}
\Psi = 
\begin{pmatrix}
\Psi_\up\\
\Psi_\down
\end{pmatrix} 
\end{equation}
and the action
\begin{equation}
 S= \int d^2x \left[  \frac{1}{2}\Psi^\top ( \partial_\tau + i \sigma_3 \partial_x + m \sigma_2 ) \Psi - g (\Psi_\up \partial_x \Psi_\up) (\Psi_\down \partial_x \Psi_\down)  \right]
.
\label{eqSM}
\end{equation}
This theory is an effective model of Majorana zero modes \cite{Hayata:2017wdw}.
The four-fermion interaction term is transformed by the Hubbard-Stratonovich method.
After the transformation, the action is
\begin{eqnarray}
&&
S_\Psi = \int d^2x \frac{1}{2}\Psi^\top \left\{ \partial_\tau + (i\sigma_3 + A) \partial_x  + m \sigma_2 \right\} \Psi
\\
&&
S_A = \int d^2x \frac{1}{2g} A^2
.
\end{eqnarray}
For lattice simulation, we introduce the Wilson term 
\begin{equation}
W\sigma_2 \equiv - \sum_\mu \frac{\partial_\mu^2}{2}\sigma_2
\label{eqW}
\end{equation}
and replace continuous derivatives by discrete differences 
\begin{eqnarray}
&&
\partial_\mu \Psi(x) \equiv \frac{1}{2} \{ \Psi(x+\hat{\mu}) - \Psi(x-\hat{\mu}) \}
\\&&
\partial_\mu^2 \Psi(x) \equiv \Psi(x+\hat{\mu}) + \Psi(x-\hat{\mu}) - 2 \Psi(x)
,
\end{eqnarray}
where $\hat{\mu}$ is the lattice vector in $\mu$ direction.
The resulting matrix is
\begin{equation}
P =
\begin{pmatrix}
\partial_\tau + D_x & -i (m+W)
 \\
i(m+W) & \partial_\tau + D_x^*
\end{pmatrix},
\end{equation}
where the matrix $D_x$ is defined by
\begin{eqnarray}
D_x \Psi_s(x) \equiv \frac{1}{2} \left[ \{i+A(x)\}\Psi_s(x+\hat{x}) - \{i+A(x-\hat{x})\}\Psi_s(x-\hat{x}) \right]
.
\end{eqnarray}
It is easy to see the matrix $P$ satisfies the Majorana positivity condition.

\begin{figure}[h]
\begin{center}
 \includegraphics[width=.7\textwidth]{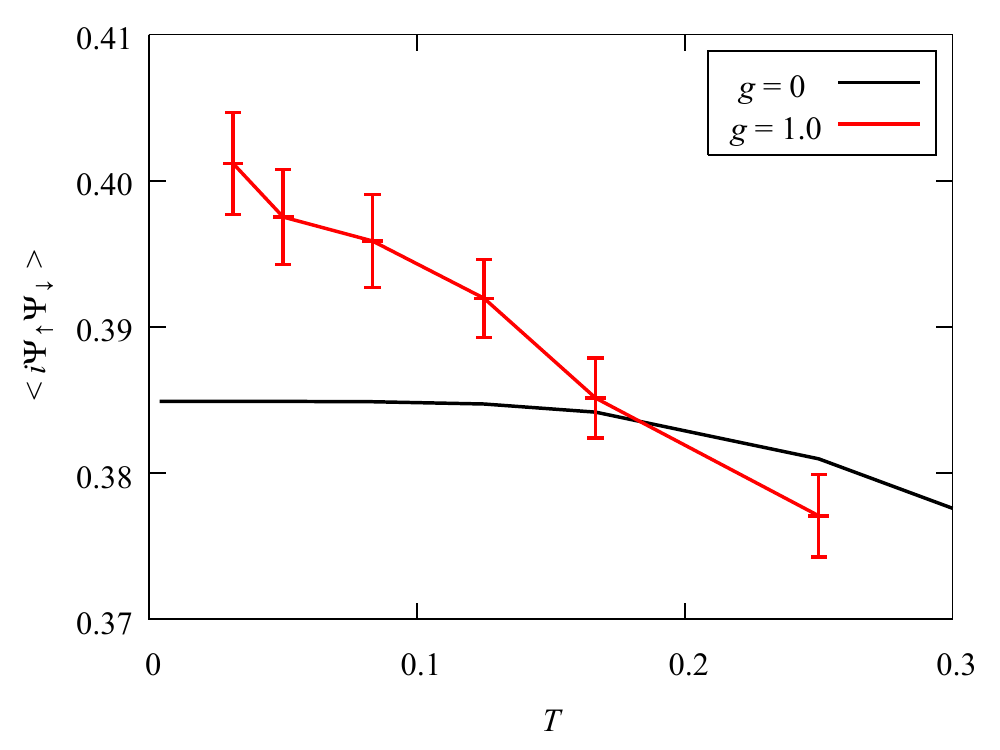}
\caption{
Pair condensate $\langle i \Psi_\up \Psi_\down \rangle$.
The simulation was done at $m=0$ and with the spatial lattice size $N_s = 32$.
\label{figC}
}
\end{center}
\end{figure}

The classical action \eqref{eqSM} has time-reversal symmetry at $m=0$, but quantum effect breaks it spontaneously.
The order parameter, the pair condensate $\langle i \Psi_\up \Psi_\down \rangle$, is shown in Fig.~\ref{figC}.
The difference between the interacting case $g=1$ and the non-interacting case $g=0$ increases at low temperature.
This suggests spontaneous breaking of time-reversal symmetry.
This phenomenon is just like spontaneous chiral symmetry breaking of the Dirac fermion.
In Fig.~\ref{figC}, we have nonzero condensate even in the non-interacting case because of the Wilson term \eqref{eqW}, like nonzero chiral condensate of the Wilson-Dirac fermion.

\section{Dirac fermions in 1+1 dimensions}

The second example is the Dirac fermion.
Although the Majorana positivity is applicable to any Dirac fermion, we here consider the Dirac fermion theory with a Majorana-type term.
The examples of the Majorana-type term are the Cooper pair source for electromagnetic superconductivity and the diquark source for color superconductivity.
In condensed matter physics, the Cooper pair source can be physically generated by the proximity effect \cite{Hasan:2010xy}.

In 1+1 dimensions, we have the two-component Dirac fermion
\begin{equation}
\psi = 
\begin{pmatrix}
\psi_\up\\
\psi_\down
\end{pmatrix} 
\end{equation}
and the charge conjugation $\psi_c$.
Let us consider the action including the source term of the Cooper pair,
\begin{equation}
 S = \int d^2x \left[ \overline{\psi} (\sigma_2 \partial_\tau + \sigma_1 \partial_x +  m) \psi - \frac{\Delta}{2} \left( \overline{\psi}_c \psi + \overline{\psi} \psi_c \right) + \frac{g}{2} (\psi^\dagger \psi)^2 \right]
.
\end{equation}
We use the Hubbard-Stratonovich transformation and obtain
\begin{eqnarray}
S_\Psi 
&=& \int d^2x \left[ \overline{\psi} \{\sigma_2 (\partial_\tau + iA) + \sigma_1 \partial_x + m\} \psi - \frac{\Delta}{2} \left( \overline{\psi}_c \psi + \overline{\psi} \psi_c \right) \right]
\nonumber\\
&=& \int d^2x \left[ \psi^\top_c C \{\sigma_2 (\partial_\tau + iA) + \sigma_1 \partial_x + m\} \psi - \frac{\Delta}{2} \left( \psi^\top C \psi + \psi^\top_c C \psi_c \right) \right]
\\
S_A &=& \int d^2x \frac{1}{2g} A^2 
\end{eqnarray}
with the charge conjugation matrix $C = i \sigma_2$.
Transforming to the Majorana field 
\begin{eqnarray}
\psi &=& \frac{1}{\sqrt{2}}(\Psi_1 + i \Psi_2)
\\
\psi_c &=& \frac{1}{\sqrt{2}}(\Psi_1 - i \Psi_2)
\end{eqnarray}
and using the basis
\begin{equation}
\Psi = 
\begin{pmatrix}
\Psi_{\up 1}\\
\Psi_{\up 2}\\
\Psi_{\down 1}\\
\Psi_{\down 2}
\end{pmatrix}
,
\end{equation}
we obtain
\begin{equation}
P =
\begin{pmatrix}
i \partial_\tau + \partial_x & - iA & m+W-\Delta & 0 \\
iA &  i \partial_\tau + \partial_x & 0 & m+W+\Delta \\
-(m+W-\Delta) & 0 &  i \partial_\tau - \partial_x & - iA \\
0 & -(m+W+\Delta) & iA & i \partial_\tau - \partial_x
\end{pmatrix}
.
\end{equation}
However, the Majorana positivity is not manifest in this basis.
We change the basis
\begin{equation}
\Psi \to \Psi' =
\begin{pmatrix}
\Psi_{\up 1} \\
\Psi_{\up 2} \\
-i \Psi_{\down 1} \\
-i \Psi_{\down 2}
\end{pmatrix}
,
\end{equation}
and obtain
\begin{equation}
P \to P' = 
\begin{pmatrix}
i \partial_\tau + \partial_x & - iA & i(m+W-\Delta) & 0 \\
iA & i \partial_\tau + \partial_x & 0 & i(m+W+\Delta) \\
-i(m+W-\Delta) & 0 & - i \partial_\tau + \partial_x & iA \\
0 & -i(m+W+\Delta) & - iA & - i \partial_\tau + \partial_x
\end{pmatrix}
.
\end{equation}
Since $m + W \ge m$, the Majorana positivity condition is satisfied when $m \ge |\Delta|$.

\begin{figure}[h]
\begin{center}
 \includegraphics[width=.7\textwidth]{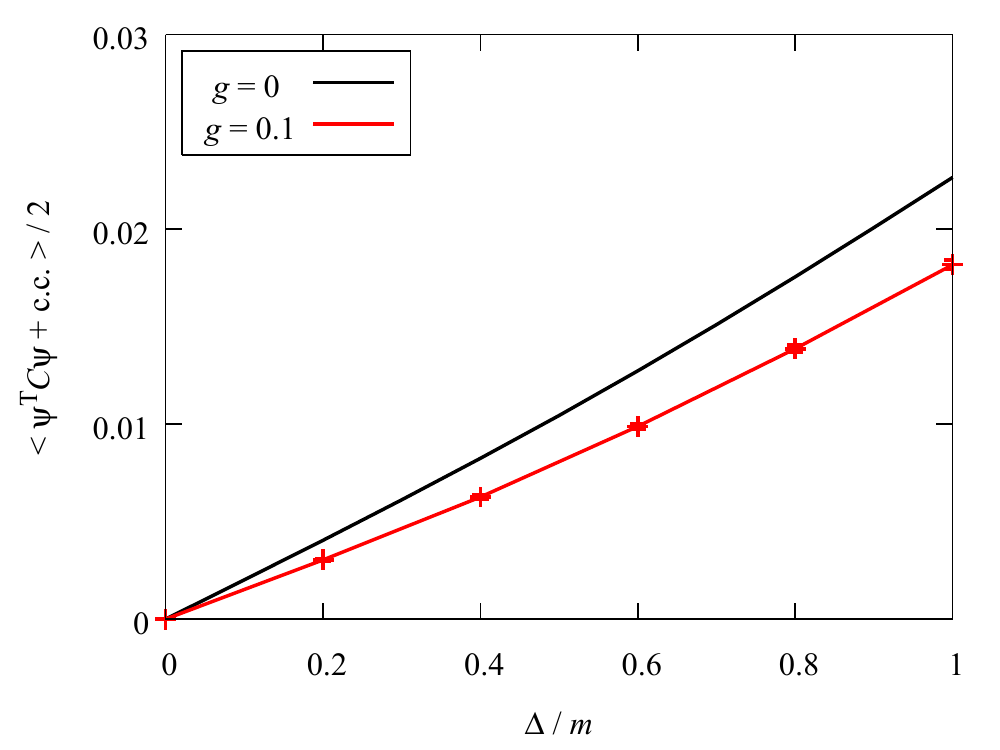}
\caption{
Cooper pair condensate $\langle  \psi^\top C \psi + \psi^\top_c C \psi_c \rangle/2$.
The simulation was done at $m=0.1$ and with the lattice volume $N_\tau N_s = 16^2$.
\label{figP}
}
\end{center}
\end{figure}

The simulation result is shown in Fig.~\ref{figP}.
Since we introduced the source term of the Cooper pair, the Cooper pair condensate $\langle  \psi^\top C \psi + \psi^\top_c C \psi_c \rangle/2$ is nonzero even in the non-interacting case $g=0$.
The condensate is suppressed in the interacting case $g=0.1$ because the repulsive interaction destroys the Cooper pair.

\section{Dirac fermions in 3+1 dimensions}

Although the (1+1)-dimensional toy models are shown above, the Majorana positivity can be applied to any dimension.
We also analyzed relativistic Dirac spinors in 3+1 dimensions.
We classified possible sign-problem-free terms and their conditions.
Known sign-problem-free terms were classified and new sign-problem-free terms were found.
The result is summarized in the tables in Ref.~\cite{Hayata:2017jdh}.

\section*{Acknowledgements}
A.~Y.~was supported by JSPS KAKENHI (Grant No.~JP15K17624).   
T.~H.~was supported by JSPS Grant-in-Aid for Scientific Research (Grant No.~JP16J02240).

\end{document}